\documentclass[aps,prl,showpacs,noshowkeys,amsmath,amssymb,amsfonts,superscriptaddress,longbibliography,reprint]{revtex4-1}
\usepackage[english]{babel}

\usepackage{graphicx}
\usepackage{bm}
\usepackage{physics}
\usepackage{mathtools}
\setcitestyle{super}
\usepackage{caption}
\usepackage{subcaption}
\DeclareCaptionLabelSeparator{bar}{~\rule[-0.4ex]{0.2ex}{1em}~}
\DeclareCaptionLabelFormat{subfor}{\textbf{#2}}
\newcommand*\bfcaption[2]{\caption[#1]{\textbf{#1.}#2}}
\usepackage{xcolor}
\definecolor{UBcolor}{HTML}{007CC1}
\usepackage[colorlinks=true,pdfnewwindow=true,linkcolor=UBcolor,citecolor=UBcolor,urlcolor=UBcolor,breaklinks=true,linktocpage]{hyperref}
\usepackage[all]{hypcap}
\usepackage[nameinlink,capitalise]{cleveref}
\begin{document}
\title{Universal scaling of active nematic turbulence}
\author{Ricard Alert}
\affiliation{Departament de F\'{i}sica de la Mat\`{e}ria Condensada, Universitat de Barcelona, Av. Diagonal 647, 08028 Barcelona, Spain}
\affiliation{Universitat de Barcelona Institute of Complex Systems (UBICS), Universitat de Barcelona, Barcelona, Spain}
\affiliation{Princeton Center for Theoretical Science, Princeton University, Princeton NJ 08544, USA}
\affiliation{Lewis-Sigler Institute for Integrative Genomics, Princeton University, Princeton NJ 08544, USA}
\author{Jean-Fran\c{c}ois Joanny}
\affiliation{ESPCI Paris, PSL Research University, 10 rue Vauquelin, 75005 Paris, France}
\affiliation{Laboratoire PhysicoChimie Curie, Institut Curie, PSL Research University - Sorbonne Universit\'{e}s, UPMC, 75005 Paris, France}
\affiliation{Coll\`{e}ge de France, 11 Place Marcellin Berthelot, 75231 Paris Cedex 05, France}
\author{Jaume Casademunt}
\affiliation{Departament de F\'{i}sica de la Mat\`{e}ria Condensada, Universitat de Barcelona, Av. Diagonal 647, 08028 Barcelona, Spain}
\affiliation{Universitat de Barcelona Institute of Complex Systems (UBICS), Universitat de Barcelona, Barcelona, Spain}
\date{\today}

\begin{abstract}
A landmark of turbulence is the emergence of universal scaling laws, such as Kolmogorov's $E(q)\sim q^{-5/3}$ scaling of the kinetic energy spectrum of inertial turbulence with the wave vector $q$. In recent years, active fluids have been shown to exhibit turbulent-like flows at low Reynolds number. However, the existence of universal scaling properties in these flows has remained unclear. To address this issue, here we propose a minimal defect-free hydrodynamic theory for two-dimensional active nematic fluids at vanishing Reynolds number. By means of large-scale simulations and analytical arguments, we show that the kinetic energy spectrum exhibits a universal scaling $E(q)\sim q^{-1}$ at long wavelengths. We find that the energy injection due to activity has a peak at a characteristic length scale, which is selected by a nonlinear mechanism. In contrast to inertial turbulence, energy is entirely dissipated at the scale where it is injected, thus precluding energy cascades. Nevertheless, the non-local character of the Stokes flow establishes long-ranged velocity correlations, which lead to the scaling behavior. We conclude that active nematic fluids define a distinct universality class of turbulence at low Reynolds number.
\end{abstract}

\maketitle

Turbulent flows exhibit universal statistical properties. Understanding how these properties emerge from the underlying governing equations is a fundamental challenge in nonequilibrium physics. In
classic inertial turbulence, energy is injected externally to drive the flows. The nonlinear advective term of the Navier-Stokes equation is responsible for destabilizing the flow and transferring energy from the scales where it is injected 
to those where it is 
dissipated. This leads to an energy cascade in the 
intermediate range of scales, where the flow acquires a scale-invariant
structure that manifests as a power-law scaling of the kinetic energy spectrum. 
Kolmogorov used these arguments in 1941 to derive the universal scaling 
exponent $-5/3$ for inertial turbulence, independent of the fluid's properties 
\cite{Kolmogorov1991,Frisch1995}. 

More recently, `elastic turbulence' was discovered in polymer solutions at low Reynolds numbers, where inertia is negligible \cite{Groisman2000}.
In the last decade, seemingly turbulent flows at low Reynolds numbers have also been 
discovered in a number of active fluids, mostly of biological origin. Examples 
include bacterial suspensions 
\cite{Dombrowski2004,Cisneros2007,Ishikawa2011,Wensink2012,Dunkel2013}, 
swarming bacteria\cite{Patteson2018} and sperm \cite{Creppy2015}, suspensions of microtubules and molecular 
motors \cite{Sanchez2012,Henkin2014,Guillamat2017,Lemma2019,Martinez-Prat2019,Tan2019}, epithelial cell 
monolayers \cite{Doostmohammadi2015,Yang2016a,Blanch-Mercader2018}, and 
suspensions of artificial self-propelled particles \cite{Nishiguchi2015,Karani2019}. These fluids 
display spontaneous flows driven by internal active stresses generated by their 
components at microscopic scales. At high activity, 
these flows become chaotic, and hence they have been referred to as active turbulence.

Hydrodynamic models for different types of active turbulence have been proposed. Motivated by suspensions of swimming bacteria, some models extend the Toner-Tu equations for 
polar flocks, thus inheriting their nonlinear alignment and polarity self-advection terms 
\cite{Wensink2012,Dunkel2013a,Dunkel2013,Slomka2015,Grossmann2014,Heidenreich2016}. Active turbulence in these models can be traced back to the same type of advective nonlinearity as in classic 
inertial turbulence: self-propulsion acts as an effective inertia that transfers energy between scales 
\cite{Bratanov2015,James2018a,Slomka2017,Slomka2018,Linkmann2019}. However, in contrast to inertial turbulence, 
these models give rise to non-monotonous flow spectra with non-universal scaling exponents, which 
depend on the values of the model parameters 
\cite{Wensink2012,Bratanov2015,Slomka2017,James2018a}.

A different class of models considers active liquid crystals, with either polar or nematic symmetry. In the 
polar case, polarity self-advection or other self-propulsion-like terms give rise to oscillatory instabilities 
that eventually lead to spatio-temporal chaos 
\cite{Wolgemuth2008,Giomi2008,Giomi2012,Bonelli2016,Ramaswamy2016,Blanch-Mercader2017c}. 
In the nematic case, however, these terms are not allowed by symmetry. Nevertheless, chaotic flows 
also appear, driven only by active stresses
\cite{Thampi2016a,Doostmohammadi2018,Fielding2011,Giomi2011,Thampi2013,
Thampi2014a,Thampi2014b,Thampi2014,Giomi2015,Thampi2015,Hemingway2016a,Doostmohammadi2017,Urzay2017,Shankar2019,Carenza2020,Coelho2019}. The balance between active stress and 
elastic nematic stress defines an intrinsic length that determines the average 
vortex size \cite{Giomi2015}. At larger scales, Giomi has proposed the existence of a scaling regime of the flow spectrum \cite{Giomi2015}. However, such a scaling has not been demonstrated yet.

Altogether, these previous studies raise the question of whether turbulence in a more classic sense, with scaling behavior and universal exponents, can exist in active fluids. Here, we show that active nematic fluids can feature turbulent flows with a universal scaling regime at large length scales. Active stresses power an instability that generates spontaneous flow, thereby injecting energy into the flow. We find that the spectrum of energy injection is broad but peaked at an intrinsic wavelength selected by the nonlinear dynamics. We also find that the injected energy is dissipated without being transferred to other scales. Therefore, the scaling regime is not sustained by an energy cascade but by the long-range hydrodynamic interactions of viscous flow.

\section*{Minimal hydrodynamic theory of active nematic fluids}

We study two-dimensional active nematic fluids at low Reynolds number. Thus, we neglect 
inertial effects, so that momentum conservation reduces to force balance:
\begin{equation} \label{eq force-balance}
0 = -\partial_\alpha P + \partial_\beta (\sigma_{\alpha\beta} + \sigma_{\alpha
\beta}^{\text{a}}).
\end{equation}
The pressure $P$ enforces the incompressibility condition $\partial_
\alpha v_\alpha = 0$ of the flow field $\vec{v}$, whereas $\sigma_{\alpha\beta}$ 
and $\sigma_{\alpha\beta}^{\text{a}}$ are the symmetric and antisymmetric parts 
of the deviatoric stress tensor, respectively. The symmetric part is given by the 
constitutive equation \cite{Kruse2005,Marchetti2013,Prost2015,Julicher2018}
\begin{equation} \label{eq stress-constitutive}
\sigma_{\alpha\beta} = 2\eta\, v_{\alpha\beta} - \zeta\,q_{\alpha\beta},
\end{equation}
where $\eta$ is the shear viscosity, $v_{\alpha\beta}=1/2(\partial_\alpha v_\beta 
+ \partial_\beta v_\alpha)$ is the symmetric part of the strain rate tensor, $\zeta$ 
is the active stress coefficient, and $q_{\alpha\beta} = n_\alpha n_\beta - 1/2\,
\delta_{\alpha\beta}$ is the nematic orientation tensor defined by the director 
field $\hat{n}$. We assume that the fluid is deep in the nematic phase so that the 
director  has a fixed modulus $|\hat{n}|=1$, and components: $n_x = \cos\theta$, 
$n_y=\sin\theta$. For a continuous director field, this constraint precludes the 
presence or generation of topological defects.
Moreover, for the sake of simplicity, we neglect the flow-alignment coupling ($\nu=0$) \cite{DeGennes-Prost}.

The antisymmetric part of the stress tensor is obtained from angular momentum 
conservation, and reads \cite{DeGennes-Prost}
\begin{equation}
\sigma_{\alpha\beta}^{\text{a}} = \frac{1}{2} (n_\alpha h_\beta - h_\alpha n_
\beta).
\end{equation}
Here, $h_\alpha = -\delta F_n/\delta n_\alpha = K\nabla^2 n_\alpha$ is the 
orientational field computed from the Frank free energy for nematic elasticity which,
in the one-constant approximation, reads \cite{DeGennes-Prost}
\begin{equation} \label{eq Frank-energy}
F_n = \frac{K}{2} \int_{\mathcal{A}} (\partial_\alpha n_\beta)(\partial_\alpha n_
\beta)\,\dd^2\vec{r} = \frac{K}{2}\int_{\mathcal{A}} |\vec{\nabla}\theta|^2\,
\dd^2\vec{r}.
\end{equation}

Finally, the dynamics of the director field reduces to
\begin{equation} \label{eq director-dynamics}
\partial_t n_\alpha + v_\beta \partial_\beta n_\alpha + \omega_{\alpha\beta} n_
\beta = \frac{1}{\gamma} h_\alpha,
\end{equation}
where $\omega_{\alpha\beta}= 1/2(\partial_\alpha v_\beta - \partial_\beta v_
\alpha)$ is the vorticity tensor, and $\gamma$ is the rotational viscosity. The left-hand 
side is the co-rotational derivative of the director field, 
whereas the orientational field on the right-hand side specifies the elastic torque 
acting on the director.

We introduce dimensionless variables by rescaling length 
by the system size $L$, 
time by the active time $\tau_a = \eta/|\zeta|$, pressure by the active stress $|
\zeta|$, and orientational field by $K/L^2$. To eliminate pressure, we take the 
curl of the force balance equation \cref{eq force-balance} and obtain a Poisson equation for the vorticity $\omega$, which we can write in terms of the stream function $\psi$ defined by $v_x=\partial_y \psi$, $v_y=-\partial_x \psi$:
\begin{subequations} \label{eq biharmonic}
\begin{align}
\nabla^2\omega & = -\nabla^4\psi = s(\vec{r},t);\label{eq vorticity}\\
s(\vec{r},t) &= \frac{1}{2}\frac{R}{A}\nabla^4 \theta + S \left[\frac{1}{2}
\left[\partial_x^2 - \partial_y^2\right] \sin 2\theta - \partial_{xy}^2 \cos 2\theta
\right]. \label{eq source}
\end{align}
\end{subequations}
This equation describes a Stokes flow stirred by a vorticity source $s(\vec{r},t)$ with two contributions. The first term 
comes from the antisymmetric stress already present in passive nematic. 
It accounts for the flow induced by the director relaxation. In contrast, the second
term accounts for the active driving.
In \cref{eq biharmonic}, we have defined three dimensionless parameters: the activity number $A \equiv  L^2/\ell_c^2$, the viscosity ratio $R\equiv \gamma/\eta$, and the sign of the active stress $S\equiv \zeta/|\zeta| =\pm 1$ for
extensile and contractile stresses, respectively. The activity number $A$ compares the system size $L$ to the active length $\ell_c=\sqrt{K/(|\zeta| R)}$ defined by the balance between active and nematic elastic stress.

Finally, in terms of the 
director angle field $\theta$ and the stream function $\psi$,
the director dynamics \cref{eq director-dynamics} reads
\begin{equation} \label{eq angle-dynamics}
\partial_t \theta + (\partial_y \psi)(\partial_x \theta) - (\partial_x\psi)(\partial_y
\theta) + \frac{1}{2}\nabla^2 \psi = \frac{1}{A} \nabla^2\theta.
\end{equation}
\Cref{eq biharmonic,eq angle-dynamics}
specify the hydrodynamics of our minimal active nematic fluid.
As shown in the \hyperref[one-parameter]
{SI, One-parameter formulation}, for a given $S=\pm 1$, the model is left with a 
single dimensionless parameter $A'=A/(2+R/2)$. Therefore, $R$ can be fixed without 
loss of generality. In the numerical simulations, we set $R=1$. Moreover, for 
$R=0$, the model takes a particularly simple form (\hyperref[one-parameter]
{SI, One-parameter formulation}).

\section*{Stationary flow patterns}

The equilibrium 
solutions of \cref{eq biharmonic,eq angle-dynamics} are uniformly-oriented quiescent states ($\psi=0,\theta=\theta_0$), with spontaneously broken rotational 
symmetry. These states are unstable to orientational fluctuations, which result in active stress fluctuations. These stresses induce flows 
that enhance orientational fluctuations, thus giving rise to the so-called spontaneous 
flow instability  
\cite{Simha2002,Voituriez2005,Marchetti2013}. The growth rate of small 
perturbations of wave-vector $\vec{q}$ forming an angle $\phi$ with the 
director $\hat{n}$ reads, in dimensional form,
\begin{equation} \label{eq growth-rate}
\Omega(\vec{q}) = \left[\frac{SA}{2}\cos 2\phi - \left(1+\frac{R}{4}\right) (qL)^2\right]\tau_r^{-1},
\end{equation}
where $\tau_r = \gamma L^2/K$. For contractile (extensile) stresses, with $S=-1$ ($S=1$), the most unstable 
perturbations are transverse (longitudinal), i.e. $\phi=\pi/2$ $(\phi=0)$, whereas
longitudinal (transverse) modulations are stable.
Hereafter, we focus on the contractile case ($S=-1$) and we fix 
$\theta_0=0$.

\begin{figure}[tb!]
\begin{center}
\includegraphics[width=\columnwidth]{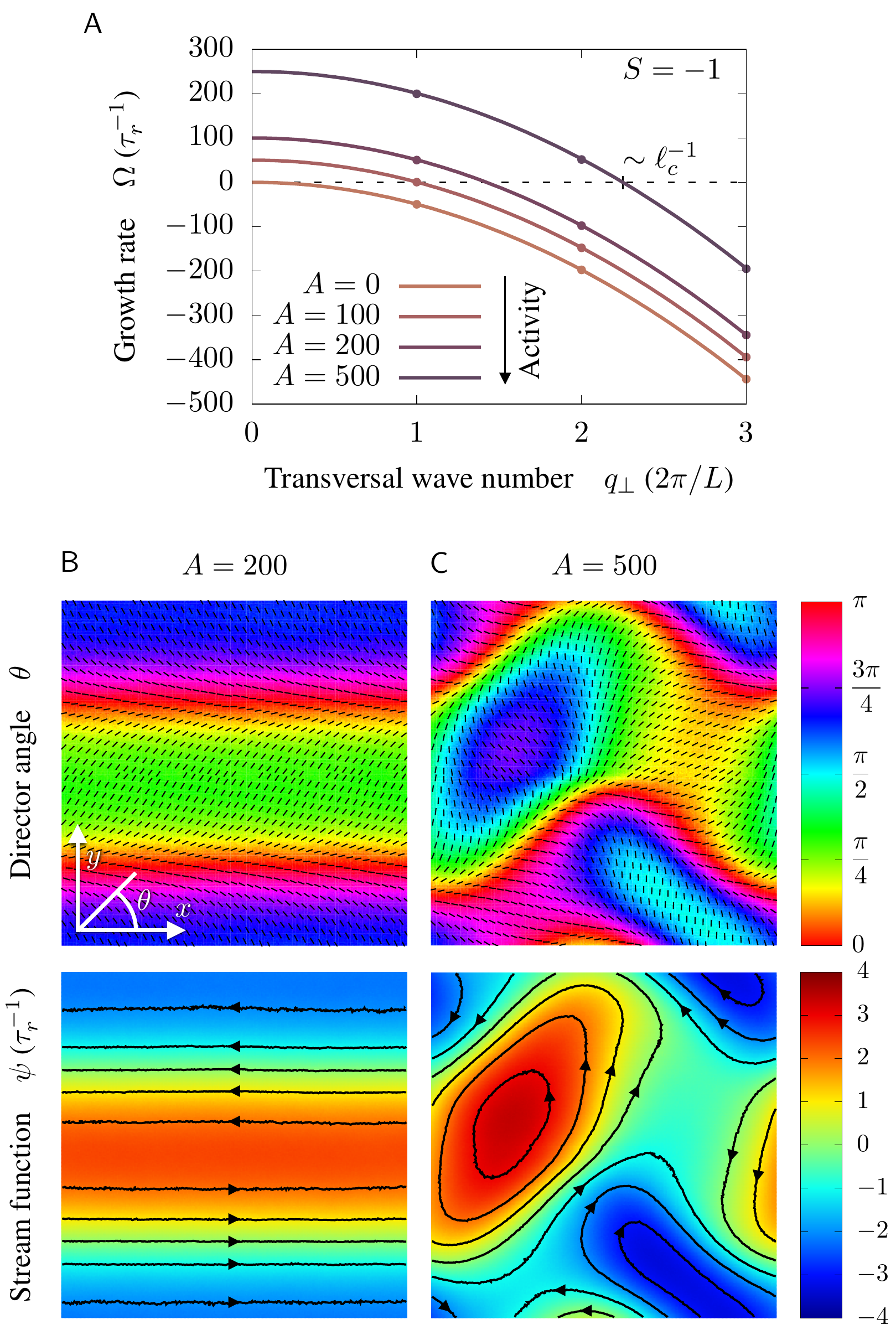}
  {\phantomsubcaption\label{Fig 1a}}
  {\phantomsubcaption\label{Fig 1b}}
  {\phantomsubcaption\label{Fig 1c}}
\bfcaption{\label{Fig 1}Stationary patterns upon the spontaneous flow instability}
{ \subref*{Fig 1a}, Growth rate \cref{eq growth-rate} of perturbations transverse to the director 
($\phi=\pi/2$) for a contractile system ($S=-1$) with $R=1$. The critical 
wavelength $\sim \ell_c$ decreases with the activity number $A$. Points 
indicate numerical results (\hyperref[Methods]{Methods}), with discrete wave 
numbers for a system of size $L$. \subref*{Fig 1b}, Stripe pattern
and the corresponding spontaneous shear 
flow for $A=200$.
\subref*{Fig 1c}, Stable vortex pattern at higher 
activity number $A=500$. Small bars indicate the director (top). Lines with 
arrows indicate streamlines (bottom).}
\end{center}
\end{figure}

The critical 
wavelength in the unstable direction, $\lambda_c = 2\pi \ell_c[2+R/2]^{1/2}$, decreases with activity. Therefore,
the uniform state becomes unstable when $\lambda_c<L$, i.e. for $A>A_c=4\pi^2(2+R/
2)$, with $A_c\approx 100$ for $R=1$ (\cref{Fig 1a}). Right past the instability threshold, only the 
longest-wavelength mode is unstable, and hence the system 
evolves into a stripe pattern of wavelength $L$ with a 
spontaneous shear along the most unstable direction (\cref{Fig 1b}).
With increasing activity, the amplitude of the pattern increases, and the domain walls, where the flow concentrates, become thinner (\hyperref[stripes]{SI, One-dimensional stripe patterns}).

At higher activity, the striped pattern 
undergoes a zig-zag instability that breaks translational invariance along the $\hat{x}$ direction. The stripes become increasingly undulated and break up into vortices
(\cref{Fig 1c}, \hyperref[movie1]{SI Movie 1}). For these vortex patterns, reflection symmetry ($\theta 
\rightarrow -\theta$, $y \rightarrow -y$, $\psi \rightarrow -\psi$) is spontaneously broken. Therefore, a pattern of vortices with the opposite orientation and vorticity is a degenerate solution. At high activity, vortex lattice solutions also exist but they are unstable (\hyperref[vortex-patterns]{SI, Vortex patterns}).
All these patterns satisfy the condition $\psi=2\theta/A + c$, with $c$ constant, such that the director angle remains constant along streamlines 
(\cref{Fig 1c}, see \hyperref[vortex-patterns]{SI, Vortex patterns}).

\begin{figure}[tb]
\begin{center}
\includegraphics[width=0.65\columnwidth]{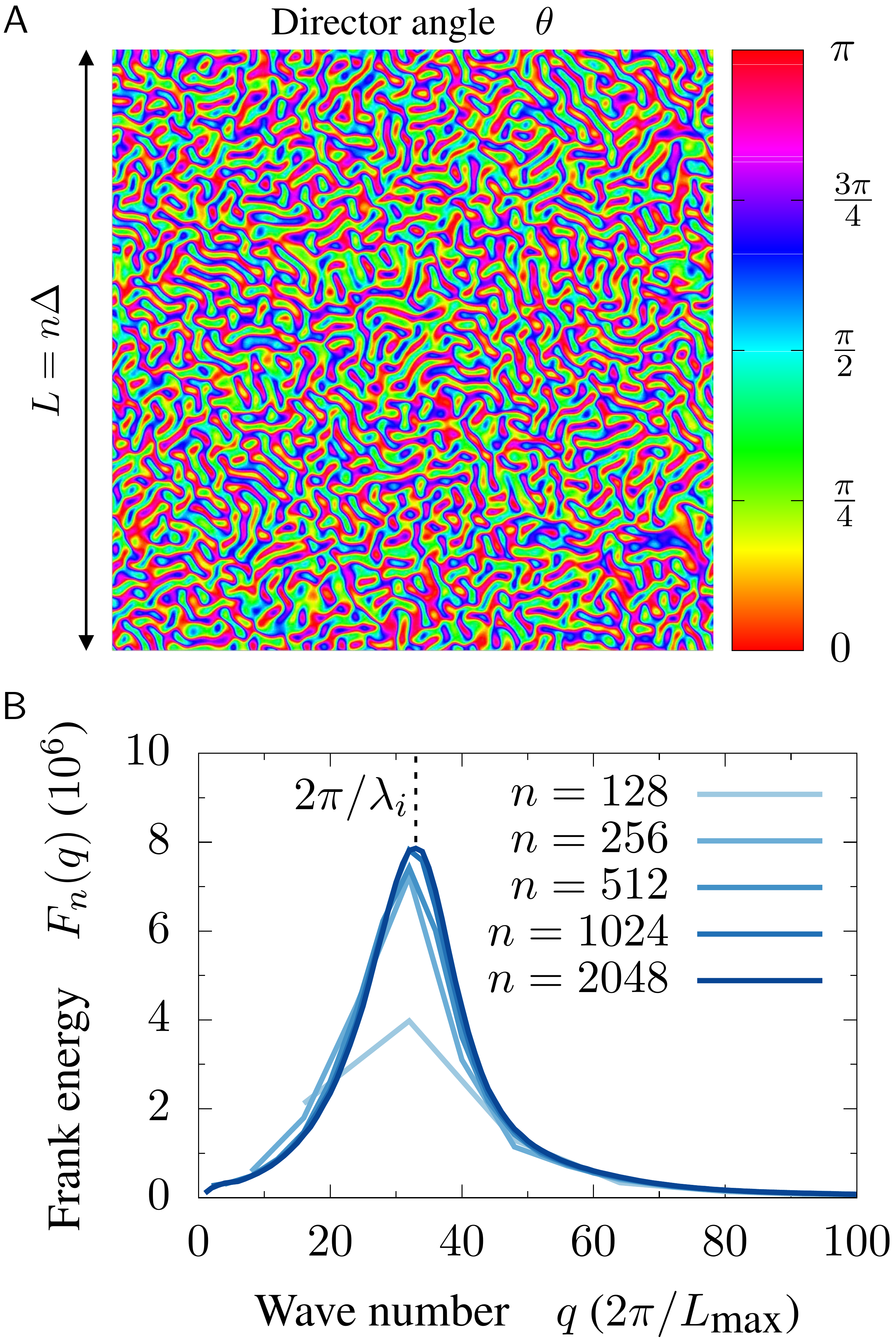}
  {\phantomsubcaption\label{Fig 2a}}
  {\phantomsubcaption\label{Fig 2b}}
\bfcaption{\label{Fig 2}Disordered pattern of director domains with a characteristic wavelength}{ \subref*{Fig 2a}, Snapshot of the angle field for a system size of $2048\times 2048$ grid points at activity number $A_{\text{max}}=3.2\cdot 10^5$. \subref*{Fig 2b}, The spectrum of the Frank elastic energy \cref{eq Frank-energy} (\cref{eq Frank-energy-spectrum}, units of $K/L_{\text{max}}$) is peaked at an intrinsic wavelength $\lambda_i\sim \ell_c$ independent of system size. Here, $\ell_c$ is held fixed ($\ell_c = L_{\text{max}}/\sqrt{A_{\text{max}}}\approx L_{\text{max}}/566$), such that the activity number $A= L^2/\ell_c^2$ increases with system size $L=n\Delta$. The wave number is rescaled by the largest system size $L_{\text{max}}$, such that the axis shows the mode number in the largest system ($n=2048$).}
\end{center}
\end{figure}

\section*{Route to turbulence and nonlinear wavelength selection}

The vortex patterns remain stable up to values of $A\sim 1500$. For larger activities, we find that
the system shows signatures of excitable dynamics 
\cite{Giomi2011,Giomi2014a}, whereby long transients with slow dynamics are 
interspersed with rapid rearrangements into a degenerate orientation of the 
pattern. Further increasing the activity, the pattern becomes increasingly 
disordered (\cref{Fig 2a}) and exhibits persistent dynamics suggestive of spatiotemporal chaos (\hyperref[movie2]{SI Movie 2}).
Overall, this sequence of dynamical patterns can be seen as a route from laminar to turbulent flow.

In the disordered chaotic patterns, the spectrum of Frank elastic energy (\cref{eq Frank-energy,eq Frank-energy-spectrum})
features a peak at a wavelength independent of system size (\cref{Fig 2b}).
However, the linear dynamics of the spontaneous-flow instability does not select any intrinsic wavelength but only the direction of the most unstable modes (\cref{eq growth-rate,Fig 1a}). Therefore, the intrinsic wavelength $\lambda_i\sim \ell_c$ of the patterns (\cref{Fig 2}) must be selected by the nonlinear dynamics of the director field (\cref{eq director-dynamics}). In fact, the nonlinear selection mechanism is based on the dynamics of evolution toward the turbulent state. For example, upon a quench from the uniform state $\theta_0=0$ to the highly turbulent regime ($A\gg 5000$), the system splits into two quiescent domains of uniform orientation $\theta=\pm\pi/2$ separated by walls of thickness $\sim \ell_c\ll L$ (\hyperref[stripes]{SI, One-dimensional stripe patterns}). The uniform domains are unstable in the perpendicular direction, and hence they also split. This process continues sequentially until the domain size is comparable to the wall thickness $\sim \ell_c$. Thereby, this transient instability cascade ends up restoring global rotational invariance and selecting a wavelength for the flow pattern. A sequential process similar to this instability cascade was recently observed experimentally \cite{Martinez-Prat2019}.

\section*{Spectral energy balance}

The instability cascade whereby the system becomes turbulent entails a transfer of energy between scales. However, beyond this transient effect, does a \emph{stationary} transfer of energy exist in active nematic turbulence? To probe possible energy fluxes across scales, we perform a spectral analysis of energy balance in the stationary turbulent regime. The energy of nematic liquid crystals includes not only the kinetic energy of the flow but also the Frank elastic energy of the director field, $F_n$. For vanishing Reynolds number, the kinetic contribution vanishes. Thus, the rate of change of the average energy, which vanishes in a statistically stationary state, can be expressed in Fourier space as (\hyperref[derivation-spectral-energy-balance]{SI, Derivation of the spectral energy balance})
\begin{equation} \label{eq power-balance}
\dot{F}_n(q) = - D_s(q) - D_r(q) + I(q) + T(q) = 0.
\end{equation}
Here, we have separated four contributions: the shear viscous dissipation rate of the flow, $D_s(q)$; the rotational 
viscous dissipation rate of the director field, $D_r(q)$; the power injected by the active stress,
$I(q)$; and the 
power $T(q)$ transferred from other scales into mode $q$, which arises 
from the advection of the director field. 
The explicit form of these contributions in both real and momentum space are given in the \hyperref[derivation-spectral-energy-balance]{SI, Derivation of the spectral energy balance}. In the following, we analyze the spectra of the contributions in \cref{eq power-balance} (\cref{eq power-spectra}).

\begin{figure}[tb]
\begin{center}
\includegraphics[width=0.75\columnwidth]{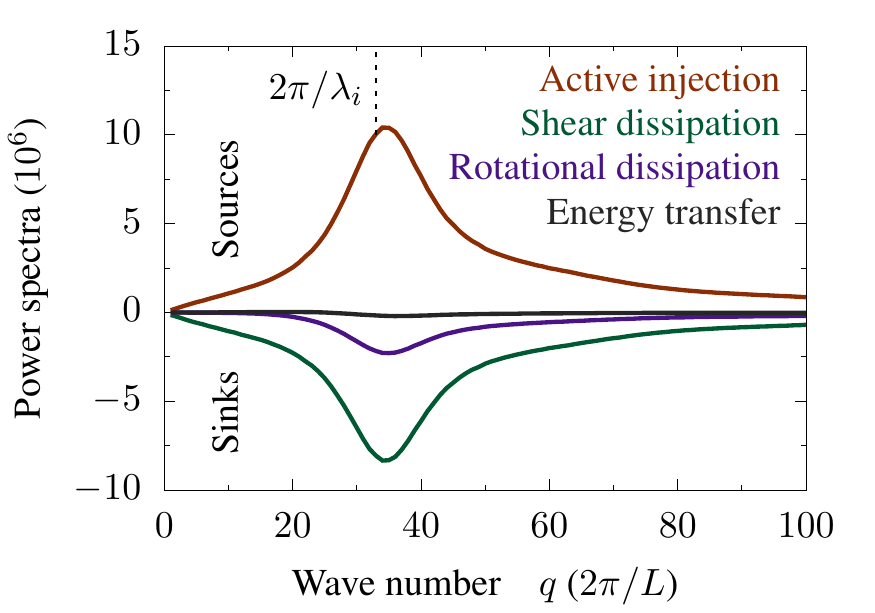}
\bfcaption{\label{Fig 3}Spectra of the four contributions to the energy balance \cref{eq power-balance} (\cref{eq power-spectra})}{ These results are for the largest system size with $2048\times 2048$ grid points at activity number $A = 3.2\cdot 10^5$. \cref{Fig S2} shows results for smaller systems. The active energy injection is entirely balanced by dissipation at every scale; no energy is transferred between scales. The energy injection rate is maximal at the selected wavelength $\lambda_i$.}
\end{center}
\end{figure}

In contrast to inertial turbulence, the energy injection is not controlled externally but it is a self-organized process. We find that the active power $I$ has a probability distribution with a positive average, meaning that active stress yields a net injection of energy into the flow (\cref{Fig S1}).
In stationary conditions, the spectral distribution of active energy injection is broad, with a maximum
at the selected wavelength $\lambda_i$ (\cref{Fig 3}), where the stored elastic energy is also maximal (\cref{Fig 2b}).

Also in contrast to inertial turbulence,
the power injection spectrum is balanced by the sum of the dissipation rate spectra locally in Fourier space (\cref{Fig 3}). This means that the energy injected at a given scale is entirely dissipated at that same scale. Therefore, there is no energy transfer 
between scales, and hence no energy cascade. Indeed, using symmetry arguments, we show that 
the energy transfer term
vanishes for all $q$, $T(\vec{q}) = 0$ (\hyperref[transfer]{SI, Absence of energy transfer between scales}), which we verify numerically (\cref{Fig 3}).

\section*{Universal scaling}

\begin{figure*}[tb]
\begin{center}
\includegraphics[width=\textwidth]{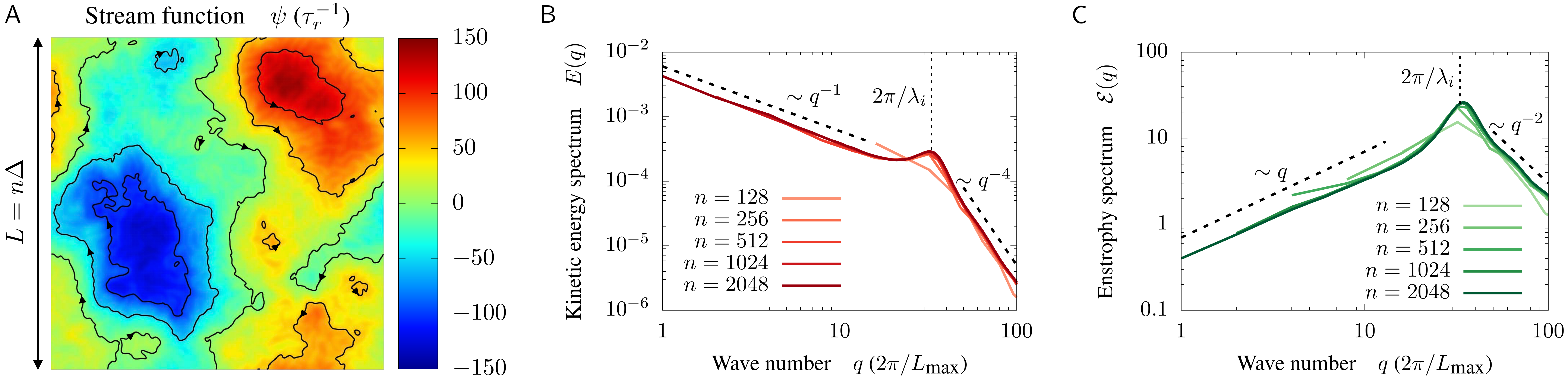}
  {\phantomsubcaption\label{Fig 4a}}
  {\phantomsubcaption\label{Fig 4b}}
  {\phantomsubcaption\label{Fig 4c}}
\bfcaption{\label{Fig 4}Universal scaling of the flow spectra at large scales}{ \subref*{Fig 4a}, Snapshot of the stream function field for a system size of $2048\times 2048$ grid points at activity number $A_{\text{max}} = 3.2 \cdot 10^5$. The small background ripples correspond to coherent vortices of typical size given by the scale of maximal energy injection $\lambda_i$. However, large-scale correlated flow patches also form due to the non-local character of viscous flow. Lines with arrows indicate a few streamlines that highlight these large-scale circulations. \subref*{Fig 4b}-\subref*{Fig 4c}, The spectra of kinetic energy and enstrophy \cref{eq kinetic-energy-enstrophy} (\cref{eq kinetic-energy-spectrum,eq enstrophy-spectrum}) exhibit a peak at the maximal injection scale $\lambda_i$. At smaller and, importantly, also at larger scales, the spectra feature distinctive universal scaling regimes. As in \cref{Fig 2b}, $\ell_c$ is held fixed ($\ell_c = L_{\text{max}}/\sqrt{A_{\text{max}}}\approx L_{\text{max}}/566$), such that the activity number $A= L^2/\ell_c^2$ increases with system size $L=n\Delta$. The wave number is rescaled by the largest system size $L_{\text{max}}$, such that the axis shows the mode number in the largest system ($n=2048$).}
\end{center}
\end{figure*}

Finally, to study the structure of the turbulent flow,
we analyze the spectra of the kinetic energy per mass density, $E$, and of the enstrophy $\mathcal{E}$:
\begin{equation} \label{eq kinetic-energy-enstrophy}
E=\frac{1}{2} \int_{\mathcal{A}} v^2\,\dd^2\vec{r},\qquad \mathcal{E} = \int_{\mathcal{A}} \omega^2\,\dd^2\vec{r}.
\end{equation}
Hereafter, we call $E$ the kinetic energy. The prominent injection of energy at the selected wavelength $\lambda_i$ gives rise to vortices of that 
typical size (the small ripples in \cref{Fig 4a}). Accordingly, the kinetic energy and the 
enstrophy spectra (\cref{eq kinetic-energy-spectrum,eq enstrophy-spectrum}) exhibit a peak at $\lambda_i$ (\cref{Fig 4b,Fig 4c}). At wavelengths smaller than this typical vortex size, 
these spectra respectively scale as $E(q)\sim q^{-4}$ and $\mathcal{E}(q) = 2q^2 E(q) \sim q^{-2}$, in 
agreement with the numerical results and mean-field predictions by Giomi \cite{Giomi2015}. These scalings characterize the internal structure of the vortices.

At scales larger than $\lambda_i$, the system develops large patches of non-coherent but correlated flow with net circulation (large-scale structures in \cref{Fig 4a}), each of which encompasses many of the smaller coherent vortices (\hyperref[movie3]{SI Movie 3}). These large-scale structures emerge from
the non-local character of Stokes flow.
Thereby, local fluctuations of the vorticity source $s$ instantaneously propagate
to the whole system through the long-range kernel of \cref{eq vorticity}. As a result, the flow field builds up long-range correlations, and hence the spectrum $E(q)$ should display scale invariance at large length scales.

To extract the scaling exponent from the governing equations,
we first analyze the spectrum of the vorticity source $s(\vec{r},t)$, which is ultimately determined by the director field $\theta(\vec{r},t)$ (\cref{eq source}). The fact that the elastic energy spectrum is peaked and does not exhibit scaling suggests that orientational fluctuations
have a finite correlation length. Therefore,
we expect $\langle |\tilde{\theta}_{\vec{q}}|^2 \rangle \sim q^0$ for $q\rightarrow 0$. Consequently, from \cref{eq source}, $\langle |\tilde{s}_{\vec{q}}|^2\rangle \sim q^4$. Hence, the vorticity source spectrum (\cref{eq vorticity-source-spectrum}) should scale as $\mathcal{S}(q) \sim q\, \langle |\tilde{s}_{\vec{q}}|^2\rangle \sim q^5$, which we verified in our two-dimensional simulations (\cref{Fig S3}).
In turn, in Fourier space, \cref{eq vorticity} reads $-q^2 \tilde{\omega}_{\vec{q}} = \tilde{s}_{\vec{q}}$, and hence $\langle |\tilde{\omega}_{\vec{q}}|^2\rangle \sim q^0$ and $\langle |\tilde{\vec{v}}_{\vec{q}}|^2\rangle =  q^{-2} \langle |\tilde{\omega}_{\vec{q}}|^2\rangle \sim q^{-2}$. Thus, we find that the kinetic energy and enstrophy spectra 
scale as $E(q)\sim q\, \langle |\tilde{\vec{v}}_{\vec{q}}|^2\rangle \sim q^{-1}$ and $\mathcal{E}(q)\sim q\, \langle |\tilde{\omega}_{\vec{q}}|^2\rangle \sim q$, which we also verify numerically
(\cref{Fig 4b,Fig 4c}).

\section*{Discussion and conclusions}

In summary, we have introduced a minimal hydrodynamic theory of an active nematic fluid at zero Reynolds number. Based on this theory, we have shown that active fluids can exhibit turbulent flows with universal scaling properties.

To maximize analytical insight and simulation power, we ignored topological defects and the flow-alignment coupling. At scales comparable to the 
intrinsic active length, the creation and annihilation of topological defects is strongly coupled to the vortex dynamics 
\cite{Giomi2015,Giomi2014a,Thampi2014b}. Nevertheless, in agreement with \cite{Giomi2015}, our results show that defects are not essential to understand the large-scale statistics of 
active turbulence. Moreover, including the flow-alignment coupling does not qualitatively modify the spontaneous flow instability \cite{Voituriez2005} that drives the turbulence. We expect that this coupling can only modify non-universal features such as the structure of vortex 
patterns and the transition to turbulence, but not the short- or long-range character of correlations. Hence, we believe that our minimal description captures the essential ingredients that determine the universal scaling properties of active nematic turbulence.

In order to understand the route to turbulence, we first studied the emergence of both stationary and turbulent flow patterns. In active nematic fluids, non-uniform director fields generate active stresses that drive spontaneous flows which, in turn, couple to the director field. At moderate values of the dimensionless activity number $A=(L/\ell_c)^2$, this unstable feedback gives rise to stable stationary patterns of orientation domains and flow vortices (\cref{Fig 1}) of increasing complexity \cite{Doostmohammadi2017,Shendruk2017}. At higher activity, the system undergoes a transition to turbulence. The turbulent state is characterized by a single characteristic length $\lambda_i\sim \ell_c$ (\cref{Fig 2}). We showed that the wavelength selection mechanism is inherently nonlinear. In contrast, in the active turbulence of generalized flocking models, the vortex size is selected by a linear instability \cite{Wensink2012,Dunkel2013,Dunkel2013a,Grossmann2014,Bratanov2015,Slomka2015,Heidenreich2016,Slomka2017a,James2018a}.

To look for universal features, we studied the large-scale statistical properties of the turbulent flows. In particular, we derived the spectral energy balance in the turbulent regime. We showed numerically that energy injection by the active processes spans all scales, and it is maximal at the selected wavelength $\lambda_i$ (\cref{Fig 3}). At vanishing Reynolds, the injected energy cannot be transferred to other scales by momentum advection as in inertial turbulence \cite{Urzay2017}. Here, we showed that, even in the presence of advection of the director field, our minimal active nematic fluid does not exhibit energy transfer between scales in the stationary turbulent state. The absence of energy transfer is possible because, in active fluids, the spectrum of energy injection is not externally fixed but results from the feedback between the director and the flow fields. As a result, the system self-organizes into a state in which energy injection is exactly balanced by dissipation at each scale. Further research is needed to determine whether flow alignment or topological defects can lead to stationary energy transfer between scales.

Finally, we studied the spectra of kinetic energy and enstrophy of the turbulent flows. Active stresses generate vortices with a characteristic scale. Hence, at the scale $\lambda_i$ of maximal active injection, the flow spectra feature at peak that reflects the underlying pattern of vortices, in agreement with previous works \cite{Giomi2015,Urzay2017}. Based on similar observations, and using mean-field arguments, Giomi predicted that the spectrum of kinetic energy would have a $E(q)\sim q^{-1}$ scaling regime at large scales. However, he could not access sufficiently large scales to verify this prediction in simulations \cite{Giomi2015}. Here, we leveraged our minimal approach to reach very large-scale simulations that allowed us to conclusively demonstrate this scaling regime (\cref{Fig 4}). Moreover, we provide an analytical understanding of the origin of the scaling and of the universal character of its exponent. In the absence of inertia, hydrodynamic interactions are long-ranged. Thereby, local stresses generate not only coherent vortices of a characteristic size but also large-scale correlated flows. These flows span a range of scales only limited by the system size, as manifest in the scaling behavior. We derived the scaling laws from the governing equations simply by assuming that the director field has a finite correlation length. Provided that this condition holds, we predict $E(q)\sim q^{-1}$.

In other words, we conclude that, via the long-range hydrodynamic interactions of viscous flow, active processes are able to maintain large-scale modes out of equilibrium and enforce power-law scaling. In this sense\cite{Falkovich2008}, chaotic flows in active nematic fluids can be called turbulent. These flows form a distinct class of turbulence at vanishing Reynolds number, in which (i) energy is injected at all scales in a self-organized way, and (ii) the flow exhibits universal scale invariance at large scales. In addition, active nematic turbulence can exist without any stationary transfer of energy across scales. This type of active turbulence is thus different from that exhibited by flocking models, which display advective energy cascades \cite{Bratanov2015,James2018a,Slomka2017,Slomka2018,Linkmann2019} and parameter-dependent scaling exponents \cite{Wensink2012,Bratanov2015,James2018a,Slomka2017}. Looking forward, we expect that our findings can be tested in large-scale experimental realizations of active nematics.

\section*{Acknowledgments}
\vskip-0.4cm
We thank Jacques Prost for discussions. R.A. thanks Anna Frishman for discussions. R.A. acknowledges support from Fundaci\'{o} ``La Caixa'' and from the Human Frontiers of Science Program (LT000475/2018-C). R.A. thanks Jacques Prost and acknowledges The Company of Biologists (Development Travelling Fellowship DEVTF-151206) and Fundaci\'{o} Universit\`{a}ria Agust\'{i} Pedro i Pons for supporting visits to Institut Curie. J.C. and R.A. acknowledge financial support by MINECO under project FIS2016-78507-C2-2-P and Generalitat de Catalunya under project 2017-SGR-1061. J.C. and J-F.J. acknowledge support from the Labex Celtisphybio ANR-10-LABX-0038 part of the Idex PSL.

\section*{Author contributions}
\vskip-0.4cm
J.C. conceived the research. R.A. and J.C. performed analytical calculations. R.A. 
performed the simulations. All authors designed the research and interpreted the results. All authors 
wrote the paper.

\section*{Competing interests}
\vskip-0.4cm
The authors declare no competing interests.

\bibliography{Active_turbulence}

\onecolumngrid 

\pagebreak

\twocolumngrid

\section*{Methods} \label{Methods}

\subsection*{Numerical scheme} 

Here, we describe the implementation of the numerical integration of our hydrodynamic equations, \cref{eq biharmonic,eq angle-dynamics}. We implement a hybrid numerical scheme that combines a spectral method for the time-independent force balance equation \cref{eq biharmonic} with a generalized version of the Alternating-Direction Implicit (ADI) algorithm \cite{Press1992} for the time evolution of the director dynamics \cref{eq angle-dynamics}. To account for fluctuations, we supplement \cref{eq angle-dynamics} with a Gaussian white noise field with $\langle \xi(\vec{r},t)\rangle =0$ and $\langle \xi(\vec{r},t)\xi(\vec{r}\,',t')\rangle = 2D\delta(\vec{r}-\vec{r}\,')\delta(t-t')$, which we implement by means of a standard stochastic algorithm \cite{Seesselberg1993}. We discretize the fields on a grid of $n\times n$ points. We keep a constant grid spacing $\Delta$, and we vary $n$ to change the system size $L=n\Delta$.

At each time step, the scheme computes the numerical Fourier transforms of the director angle field $\theta(\vec{r},t)$ and of the nonlinear terms on the right-hand side of \cref{eq source}. We apply the $2/3$ rule to prevent aliasing in the Fourier components \cite{Press1992}. From them, we compute the Fourier components of the stream function field $\psi(\vec{r},t)$ from the spectral decomposition of \cref{eq biharmonic}. In dimensionless variables, it reads
\begin{equation} \label{eq Fourier-stream-function}
\tilde{\psi}_{\vec{q}} = -\frac{R}{2A} \tilde{\theta}_{\vec{q}} + \frac{S}{q^4 +\epsilon} \left[\frac{q_x^2-q_y^2}{2}\mathcal{F}[\sin 2\theta]_{\vec{q}} - q_x q_y \mathcal{F}[\cos 2\theta]_{\vec{q}}\right],
\end{equation}
where $\mathcal{F}[\cdot]$ indicates the Fourier transform operator, and $\epsilon = 10^{-8}$ is a numerical cut-off to avoid the divergence of the $q=0$ mode. Then, the Fourier components are transformed back to real space to update the angle field according to the stochastic version of \cref{eq angle-dynamics}. To this end, in addition to adding the noise term, we implemented two modifications of the standard ADI algorithm, which was originally designed to invert only the Laplacian operator. First, we discretize the advective terms in \cref{eq angle-dynamics} by means of centered finite differences. Second, we leverage the Sherman-Morrison formula to impose periodic boundary conditions \cite{Press1992}.

\subsection*{Numerical tests}

Numerical results were benchmarked against analytical results. In particular, we checked the growth rate, \cref{eq growth-rate}, as well as the saturation angle of the transversal stationary patterns, \cref{eq saturation-angle}. The integral in \cref{eq saturation-angle} was numerically approximated by summing 10000 terms of the associated Legendre polynomial \cite{Chaichian2012}:
\begin{equation}
\sqrt{\frac{2A}{4+R}} = 2\pi \sum_{k=0}^\infty \left[\frac{(2k-1)!!}{2^k \,k!}\right]^2 \sin^{2k} \frac{\theta_s}{2}.
\end{equation}

\subsection*{Numerical details}

All numerical integrations have been performed for contractile systems ($S=-1$) with $R\equiv\gamma/\eta=1$. The amplitude of the angular noise is set to $D=5\cdot 10^{-4}$ $L^2/\tau_r$. In all cases, the initial condition was a quiescent state with uniform director along the $\hat{x}$ axis, namely $\theta_0=0$. The integration time step is reduced as the number of grid points is increased (\cref{table time-step}).

\begin{table}[bt]
\begin{center}
\begin{tabular}{c|ccccc}
$n$&$128$&$256$&$512$&$1024$&$2048$\\
$\Delta t$ $(\tau_r)$&$10^{-4}$&$10^{-5}$&$5\cdot 10^{-6}$&$10^{-6}$&$2\cdot 10^{-7}$
\end{tabular}
\end{center}
\bfcaption{Integration time step $\Delta t$ for simulations with different number of grid points $n\times n$, corresponding to system sizes $L=n\Delta$}{} \label{table time-step}
\end{table}

\subsubsection*{Stationary flow patterns}

The snapshots of the stationary patterns in \cref{Fig 1b,Fig 1c} were obtained from simulations run for a time $t=0.4\tau_r$ on a grid of $256\times 256$ points.

\subsubsection*{Numerical computation of energy and power spectra}

All spectra are numerically computed by replacing the ensemble average by an average over 925 snapshots of simulations run for a time $t=0.1\tau_r$. To allow for temporal decorrelation, the snapshots are taken every $\delta t=10^{-4}\tau_r$. To allow the system to reach a statistically stationary state, the snapshots are only taken after an initial transient of $t_s=7.5\cdot 10^{-3}\tau_r$. Using these snapshots, we compute a histogram of the corresponding spectral quantity over wave vector moduli which, for the isotropic correlations of the turbulent state (see \cref{eq isotropic-spectra}), corresponds to the angular average of the spectrum.

\section*{Data availability} \label{Data}
All the data presented this study are available upon request.

\section*{Code availability} \label{Code}
All the computer code used in this study is available upon request.

\onecolumngrid


\clearpage
\appendix

\onecolumngrid
\begin{center}
\textbf{\large Supporting Information for ``Universal scaling of active nematic turbulence''}
\end{center}

\setcounter{equation}{0}
\setcounter{figure}{0}
\renewcommand{\theequation}{S\arabic{equation}}
\renewcommand{\thefigure}{S\arabic{figure}}

\twocolumngrid

\section*{One-parameter formulation} \label{one-parameter}

In this section, we show that the model can be reformulated so that the dimensionless parameters $A$ and $R$ combine into a redefined activity number $A'$. The reformulation is based on the following
change of variables:
\begin{equation}
\psi'=\psi - \frac{2}{A}\theta. \label{eq change}
\end{equation}
In terms of the transformed vorticity $\omega'=-\nabla^2 \psi'$, \cref{eq biharmonic,eq angle-dynamics} read
\begin{subequations} \label{eq force-balance-prime}
\begin{equation}
\nabla^2 \omega' = s'(\vec{r},t); \label{eq omegaprime}
\end{equation}
\begin{equation}
s' = \frac{1}{A'}\nabla^4\theta + S \left[\frac{1}{2}\left[\partial_x^2 - \partial_y^2\right] \sin 2\theta - 
\partial_{xy}^2 \cos 2\theta\right] \label{eq sprime},
\end{equation}
\end{subequations}
\begin{equation}
\frac{Dn_\alpha}{Dt} = 0 \label{eq totalderiv}.
\end{equation}
Here, we have used the corotational derivative $Dn_\alpha/Dt = 
\partial_t n_\alpha +v'_\beta \partial_\beta n_\alpha + \omega'_{\alpha \beta} n_\beta$. In terms of the transformed stream function $\psi'$ and the director angle $\theta$, \cref{eq totalderiv} reads
\begin{equation} \label{eq angleprime}
\partial_t \theta + (\partial_y \psi')(\partial_x \theta) - (\partial_x\psi')(\partial_y\theta) + \frac{1}{2}
\nabla^2 \psi' = 0.
\end{equation}
In the transformed variables, \cref{eq force-balance-prime,eq totalderiv} depend on a single dimensionless parameter, $A'=A/(2+R/2)$. This transformed activity number corresponds to the ratio 
between the active time scale $\tau_a = \eta/|\zeta|$ and the director relaxation time scale,
taking into account both rotational dissipation and the viscous dissipation of the flow generated by the 
rotation of the director.

The transformed force balance equation, \cref{eq force-balance-prime}, is the same as the original one, \cref{eq biharmonic}, albeit for the modified activity parameter $A'$ instead of $A$. In contrast, the transformed dynamics of the director field, \cref{eq totalderiv}, takes the form of a total derivative.
This means that there is no dissipation associated to the director field, which is simply advected by the transformed velocity field and co-rotated by the transformed vorticity field. Accordingly, with respect to \cref{eq angle-dynamics}, \cref{eq angleprime} lacks the right-hand-side term, which accounts for rotational dissipation.

Finally, since \cref{eq force-balance-prime,eq totalderiv} depend on a single parameter $A'(A,R)$, we can choose the value of $R$ without loss of generality. A particularly simple case is obtained for $R=0$, which corresponds to a vanishing rotational viscosity $\gamma=0$.

\section*{One-dimensional stripe patterns} \label{stripes}

In this section, we provide further analytical details about the one-dimensional stripe patterns that form initially upon the spontaneous flow instability. These simple patterns are stable at low activity and become unstable at higher activity (\cref{Fig 1b,Fig 1c}).

For contractile stresses, the uniformly oriented state features unstable long-wavelength transverse modes ($q_y<q_c$, \cref{Fig 1a}) and stable longitudinal modes ($q_x$). Therefore, the emerging pattern is transversal, and it preserves translational invariance along the direction of the initial uniform orientation ($\hat{x}$ axis).

\subsection*{Dynamics. The sine-Gordon equation} \label{stripes-dynamics}

In this situation, \cref{eq biharmonic,eq angle-dynamics} can be combined and reduce to
\begin{equation} \label{eq sinegordon}
2 \partial_t \theta = \frac{1}{A'}\partial_y^2 \theta + \frac{1}{2}\sin{2\theta},
\end{equation}
which is the overdamped sine-Gordon equation. This equation can also be written in an explicitly variational form:
\begin{subequations}
\begin{equation}
2 \partial_t \theta = - \frac{\delta {\mathcal{L}}[\theta(y)]}{\delta \theta(y)};
\end{equation}
\begin{equation}
{\mathcal{L}}[\theta(y)] = \frac{1}{2} \int \left( \frac{1}{A'} (\partial_{y'}\theta)^2 - \sin^2\theta  \right) \dd y',
\end{equation}
\end{subequations}
where the Lyapunov functional $\mathcal{L}$ is minimized by the evolution. This problem maps onto that of a passive nematic liquid crystal in the twist geometry \cite{DeGennes-Prost}, with $\mathcal{L}$ corresponding to the free energy, and the activity playing the role of an external magnetic field. At a critical value of the external field, the passive liquid crystal undergoes the Fréedericksz transition. Similarly, the active system undergoes the spontaneous flow instability at a critical value of the activity. However, there is a fundamental difference between both transitions: whereas isotropy is externally broken by the magnetic field in the passive case, it is spontaneously broken in the active case. This means that the same dynamics of \cref{eq sinegordon} applies to uniform domains oriented in any direction, which will also be unstable to their own transverse fluctuations. This fact is precisely what gives rise to the instability cascade that characterizes the route to turbulence in our model, as described in the main text.

\subsection*{Stationary patterns. Mapping to the simple pendulum} \label{stationary-stripes}

In the stationary state, \cref{eq sinegordon} reduces to
\begin{equation}
\frac{\dd^2\theta}{\dd y^2} = -\frac{A'}{2}\sin 2\theta. \label{eq theta-stripe}
\end{equation}
This equation describes the director angle profile along the stripe pattern. Since the fastest-growing mode is the longest-wavelength one (\cref{Fig 1a}), the wavelength of the pattern is the system size $L$ (\cref{Fig 1b}).

To solve \cref{eq theta-stripe}, it is useful to realize that it maps to the equation of motion of a pendulum, $\ddot{\varphi} = -g/\ell \sin\varphi$, under the following correspondences: $\varphi \leftrightarrow 2\theta$, $t\leftrightarrow y$, $g/\ell\leftrightarrow A'$. Accordingly, the wavelength of the pattern, $L$, corresponds to the pendulum's oscillation period. Respectively, the amplitude of the pattern, i.e. the saturation angle of the director profile, $\theta_s$, corresponds to the pendulum's oscillation amplitude. By virtue of this correspondence, we can use the relationship between a pendulum's period and amplitude to find the amplitude of the pattern as a function of the activity number $A'$. Hence, $\theta_s$ is implicitly given by the following complete elliptic integral of the first kind \cite{Chaichian2012}:
\begin{equation} \label{eq saturation-angle}
\frac{\sqrt{A'}}{4} = \int_0^{2\pi} \frac{\dd \alpha}{\sqrt{1-\sin^2 \theta_s \sin^2 \alpha}}.
\end{equation}
The saturation angle $\theta_s$ increases with the activity number $A'$, with $\theta_s\rightarrow \pm \pi/2$ as $A'\rightarrow \infty$.

In the high activity limit, the angle pattern would feature two domains of $\theta=\theta_s\rightarrow\pm \pi/2$ separated by walls. In a nematic liquid crystal, the angles $\pm \pi/2$ correspond to identical orientations, and the two domain walls differ by the sense of rotation of the angle across them. In the limit of large system size $L\rightarrow \infty$, the full angle profile of the stripe pattern reads
\begin{equation}
\theta(y) = \mp \theta_s \pm2 \atan \left[\exp(\sqrt{A'}(y-y_0))\right],
\end{equation}
After restoring the units, the wall thickness is $\ell'_c=\ell_c[2+R/2]^{1/2}$.

Note that we assume here that the winding number of the director field in the initial condition vanishes, and thus we exclude any solution with a finite winding number. Solutions with a finite winding number would correspond to higher-energy states where the pendulum rotates instead of oscillate.

\section*{Vortex patterns} \label{vortex-patterns}

In this section, we provide further details about the stationary vortex patterns that form at moderate activity (\cref{Fig 1c}).

\subsection*{Periodic vortex patterns} \label{vortex-lattices}

In addition to the patterns in \cref{Fig 1c}, new branches of solutions appear at activity values $A=m^2 A_c$; $m=2,3,\ldots$. These new solutions correspond to vortex lattices, i.e. periodic repetitions of the pattern in \cref{Fig 1c}, with period $L/m$ in both axes, where $L$ is the system size. We have checked numerically that these periodic vortex patterns are unstable to perturbations of wavelength larger than their period $L/m$.

\subsection*{Streamlines of stationary flow patterns} \label{streamlines-patterns}

All stationary flow patterns correspond to equilibrium solutions of the transformed equations of the one-parameter formulation, \cref{eq force-balance-prime,eq totalderiv}. Equilibrium solutions of \cref{eq force-balance-prime,eq totalderiv} are given by $s'=0$ and $\psi'=c$, where $c$ is a constant. Therefore, in the transformed variables, these solutions correspond to quiescent states in which the active stresses are exactly balanced by the nematic elastic stresses so that the vorticity source vanishes. In the original variables, these solutions correspond to stationary flows with $\psi = 2\theta/A + c$ (\cref{eq change}), for which the director remains constant along streamlines (\cref{Fig 1b,Fig 1c}).

\section*{Derivation of the spectral energy balance} \label{derivation-spectral-energy-balance}

Here, we derive the spectral energy balance equation and identify its different contributions, \cref{eq power-balance}. We do this in two steps. First, using the dynamical equations of our model, \cref{eq biharmonic,eq angle-dynamics}, we directly derive an expression for the rate of change of the Frank elastic energy \cref{eq Frank-energy}. Second, based on the nonequilibrium thermodynamics of active nematic fluids, we identify the different contributions, namely the shear and rotational dissipations, the active energy injection, and the energy transfer between scales.

\begin{figure}[tb]
\begin{center}
\includegraphics[width=0.75\columnwidth]{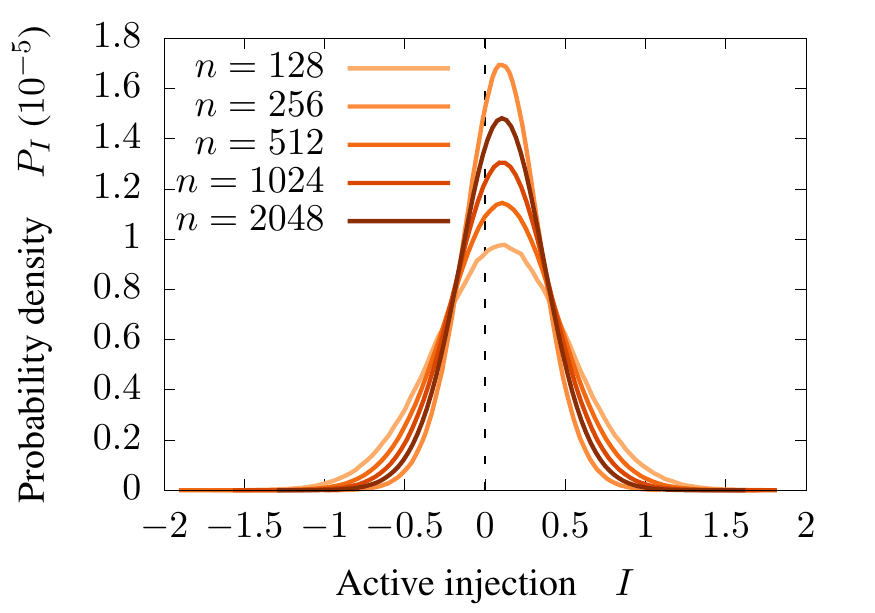}
\bfcaption{\label{Fig S1}Probability distribution of the active power $I$ for different system sizes $L=n\Delta$}{ Here, $\ell_c$ is held fixed ($\ell_c = L_{\text{max}}/\sqrt{A_{\text{max}}}\approx L_{\text{max}}/566$), such that the activity number $A= L^2/\ell_c^2$ increases with system size $L$. Negative and positive values of $I$ respectively correspond to active energy absorption and injection into the system. Therefore, on average, active stresses inject energy to drive the flows.}
\end{center}
\end{figure}

\subsection*{Direct derivation} \label{direct-derivation}

We define the spectrum of the Frank energy, $F_n(\vec{q})$, from
\begin{equation} \label{eq non-isotropic-spectra}
\frac{\langle F_n\rangle}{\mathcal{A}} \equiv \int_{\mathbb{R}^2} F_n(\vec{q})\,\dd^2\vec{q},
\end{equation}
where $\mathcal{A}=L^2$ is the system area.
From \cref{eq Frank-energy} and introducing the Fourier transform of the angle field, $
\theta(\vec{r},t)=\sum_{\vec{q}} \tilde{\theta}_{\vec{q}} \,e^{i\vec{q}\cdot\vec{r}}$, we obtain
\begin{equation}
\langle F_n\rangle = \frac{K}{2} \int_{\mathcal{A}} \langle |\vec{\nabla}\theta|^2\rangle \,\dd^2\vec{r} = 
\frac{K}{2}(2\pi)^2 L^2 \sum_{\vec{q}} q^2 \langle |\tilde{\theta}_{\vec{q}}|^2\rangle,
\end{equation}
where the sum over $\vec{q}$ runs over 
$\vec{q}=2\pi/L\, (n_x,n_y)$, with $n_x,n_y\in \mathbb{N}$. Then, in the continuum limit, we obtain
\begin{equation} \label{eq non-isotropic-Frank-spectrum}
\langle F_n\rangle = \frac{K}{2} L^4 \int_{\mathbb{R}^2} q^2 \langle |\tilde{\theta}_{\vec{q}}|^2 \rangle \,
\dd^2\vec{q},
\end{equation}
from where
\begin{equation} \label{eq Frank-spectra}
F_n(\vec{q}) = \frac{K}{2} L^2 q^2 \langle |\tilde{\theta}_{\vec{q}}|^2 \rangle.
\end{equation}

In terms of its spectrum, the rate of change of the average Frank elastic energy can be expressed as
\begin{equation}
\frac{1}{\mathcal{A}}\frac{\dd \langle F_n\rangle}{\dd t} = \int_{\mathbb{R}^2} \partial_t F_n(\vec{q})\,\dd^2 \vec{q}.
\end{equation}
Then, using \cref{eq Frank-spectra}, and in dimensionless form, we obtain
\begin{equation} \label{eq Frank-spectrum-dynamics}
\partial_t F_n(\vec{q}) = \frac{q^2}{2} \partial_t \langle |\tilde{\theta}_{\vec{q}}|^2 \rangle = q^2 \Re\left[\left\langle \tilde{\theta}^*_{\vec{q}} \,\partial_t \tilde{\theta}_{\vec{q}}\right\rangle\right].
\end{equation}
To compute the time derivative, we Fourier-transform the dynamical equation of the angle field, \cref{eq angle-dynamics}. This gives
\begin{equation} \label{eq angle-Fourier-dynamics}
\partial_t \tilde{\theta}_{\vec{q}} = q^2 \left(\frac{1}{2}\tilde{\psi}_{\vec{q}} - \frac{1}{A} \tilde{\theta}_{\vec{q}}\right) + \frac{1}{(2\pi)^2} \int_{\mathbb{R}^2} \left(\vec{k}\times\vec{q}\right)\cdot\hat{z}\, \tilde{\theta}_{\vec{k}}\,\tilde{\psi}_{\vec{q}-\vec{k}} \,\dd^2\vec{k},
\end{equation}
where the three terms correspond to director corotation, rotational dissipation, and director advection, respectively. Since the advective term is bilinear in the angle and stream function fields, it involves interactions between different Fourier modes. Introducing \cref{eq angle-Fourier-dynamics} into \cref{eq Frank-spectrum-dynamics}, we obtain
\begin{multline} \label{eq Frank-spectrum-dynamics-next}
\partial_t F_n(\vec{q}) = q^4 \left(\frac{1}{2}\Re\left[\left\langle\tilde{\psi}_{\vec{q}}\,\tilde{\theta}^*_{\vec{q}}\right\rangle\right] - \frac{1}{A} \left\langle |\tilde{\theta}_{\vec{q}}|^2 \right\rangle \right) \\
+ \frac{q^2}{(2\pi)^2} \int_{\mathbb{R}^2} \left(\vec{k}\times\vec{q}\right)\cdot\hat{z}\, \Re \left[\left\langle \tilde{\theta}_{\vec{k}}\,\tilde{\psi}_{\vec{q}-\vec{k}}\,\tilde{\theta}^*_{\vec{q}}\right\rangle\right]\,\dd^2\vec{k}.
\end{multline}
Finally, using the Fourier components of the force balance equation, \cref{eq Fourier-stream-function} with $\epsilon=0$, to replace $\tilde{\theta}^*_{\vec{q}}$ in the first term of \cref{eq Frank-spectrum-dynamics-next}, we arrive at
\begin{multline} \label{eq Frank-spectrum-dynamics-final}
\partial_t F_n(\vec{q}) = - \frac{A}{R} q^4 \left\langle |\tilde{\psi}_{\vec{q}}|^2 \right\rangle -\frac{1}{A} q^4 \left\langle |\tilde{\theta}_{\vec{q}}|^2 \right\rangle  - \frac{A}{R} \Re \left[\left\langle \tilde{s}^*_{a,\vec{q}}\,\tilde{\psi}_{\vec{q}} \right\rangle \right]\\
+ \frac{q^2}{(2\pi)^2} \int_{\mathbb{R}^2} \left(\vec{k}\times\vec{q}\right)\cdot\hat{z}\, \Re \left[\left\langle \tilde{\theta}_{\vec{k}}\,\tilde{\psi}_{\vec{q}-\vec{k}}\,\tilde{\theta}^*_{\vec{q}}\right\rangle\right]\,\dd^2\vec{k}.
\end{multline}
Here, $\tilde{s}_{a,\vec{q}}$ are the Fourier components of the active part of the vorticity source field defined in \cref{eq source}. Explicitly,
\begin{equation} \label{eq active-vorticity-source}
s_a(\vec{r},t) = S \left[\frac{1}{2}\left[\partial_x^2 - \partial_y^2\right] \sin 2\theta - \partial_{xy}^2 \cos 2\theta\right].
\end{equation}

\subsection*{Nonequilibrium thermodynamics} \label{nonequilibrium-thermodynamics}

In the following, we identify the physical meaning of the different terms in \cref{eq Frank-spectrum-dynamics-final}. Within the framework of linear nonequilibrium thermodynamics, the rate of entropy production $\dot{S}$ can be written in terms of products of fluxes and forces. For isothermal systems, $T\dot{S} = -\dot{F}$, where $F$ is the free energy. In particular, for a two-dimensional incompressible nematic fluid with fixed modulus of the director field, the rate of change of the free energy reads \cite{Kruse2005,Marchetti2013,Julicher2018,DeGennes-Prost}
\begin{equation} \label{eq dissipation-rate}
-\dot{F} = \int_{\mathcal{A}} [ \sigma_{\alpha\beta} v_{\alpha\beta} + N_\alpha h_\alpha ]\,\dd^2\vec{r}.
\end{equation}
Here, $\sigma_{\alpha\beta}$ is the symmetric part of the deviatoric stress tensor (\cref{eq stress-constitutive}), and $v_{\alpha\beta}$ is the symmetric part of the strain rate tensor. Respectively, $N_\alpha \equiv \partial_t n_\alpha + v_\beta \partial_\beta n_\alpha + \omega_{\alpha\beta} n_\beta$ is the comoving and corotational derivative of the director, which is the flux conjugated to the molecular field $h_\alpha = -\delta F_n/\delta n_\alpha$ (see \cref{eq Frank-energy}). In \cref{eq dissipation-rate}, we have not included active processes, which are usually accounted for by adding a term $r\Delta\mu$, where $r$ is the rate and $\Delta\mu$ is the chemical potential difference of the chemical reaction, such as ATP hydrolysis, that is the source of activity. Here, we describe neither the dynamics nor the energetics of this chemical reaction. Then, introducing the constitutive relations for our minimal active nematic fluid, \cref{eq stress-constitutive,eq director-dynamics} into \cref{eq dissipation-rate}, we obtain
\begin{equation} \label{eq entropy-production}
-\dot{F} = \int_{\mathcal{A}} \left[ 2\eta\, v_{\alpha\beta} v_{\alpha\beta} + \frac{1}{\gamma}\, h_\alpha h_\alpha - \zeta\, q_{\alpha\beta} v_{\alpha\beta} \right] \dd^2\vec{r}.
\end{equation}
From here, we identify the different sources of entropy production, namely viscous and rotational dissipation and active energy injection,
\begin{subequations} \label{eq power-contributions}
\begin{gather}
\begin{align}
D_s & =  \int_{\mathcal{A}} 2\eta \,v_{\alpha\beta} v_{\alpha\beta}\,\dd^2\vec{r},\\
D_r & = \int_{\mathcal{A}} \frac{1}{\gamma} \,h_\alpha h_\alpha\,\dd^2\vec{r},\\
I & = \int_{\mathcal{A}} \zeta \,q_{\alpha\beta} v_{\alpha\beta}\,\dd^2\vec{r},
\end{align}
\end{gather}
\end{subequations}
respectively. The spectra of these quantities, defined analogously to \cref{eq non-isotropic-spectra}, read, in dimensionless form
\begin{subequations} \label{eq power-spectra-contributions}
\begin{gather}
\begin{align}
D_s (\vec{q}) & =  \frac{A}{R} q^4 \left\langle |\tilde{\psi}_{\vec{q}}|^2 \right\rangle,\\
D_r (\vec{q}) & = \frac{1}{A} q^4 \left\langle |\tilde{\theta}_{\vec{q}}|^2 \right\rangle,\\
I (\vec{q}) & =  -\frac{A}{R} \Re \left[\left\langle \tilde{s}^*_{a,\vec{q}}\,\tilde{\psi}_{\vec{q}} \right\rangle \right],
\end{align}
\end{gather}
\end{subequations}
where $\Re[\cdot]$ refers to the real part, and $\tilde{s}_{a,\vec{q}}$ are the Fourier components of the active vorticity source field in \cref{eq active-vorticity-source}. Identifying these expressions in the first three terms of \cref{eq Frank-spectrum-dynamics-final}, we can rewrite that equation as
\begin{equation}
\partial_t F_n (\vec{q}) = -D_s (\vec{q}) - D_r (\vec{q}) + I(\vec{q}) + T(\vec{q}).
\end{equation}
Here, we have identified the last term in \cref{eq Frank-spectrum-dynamics-final} as the spectrum of the energy transfer between scales,
\begin{equation} \label{eq power-transfer-spectrum}
T(\vec{q}) = \frac{q^2}{(2\pi)^2} \int_{\mathbb{R}^2} \left(\vec{k}\times\vec{q}\right)\cdot\hat{z}\, \Re \left[\left\langle \tilde{\theta}_{\vec{k}}\,\tilde{\psi}_{\vec{q}-\vec{k}}\,\tilde{\theta}^*_{\vec{q}}\right\rangle\right]\,\dd^2\vec{k}.
\end{equation}
This energy transfer term does not contribute to the integrated entropy production:
\begin{equation} \label{eq integrated-transfer}
\int_{\mathbb{R}^2} T(\vec{q})\,\dd^2\vec{q} = 0.
\end{equation}

\section*{Absence of energy transfer between scales} \label{transfer}

\begin{figure}[tb]
\begin{center}
\includegraphics[width=0.75\columnwidth]{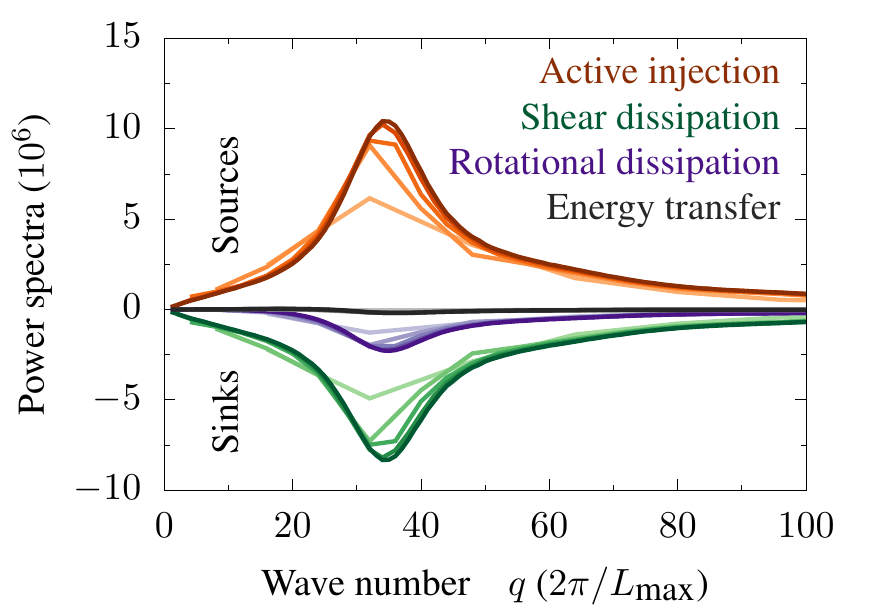}
\bfcaption{\label{Fig S2}Spectra of the four contributions to the energy balance \cref{eq power-balance} (\cref{eq power-spectra}) for different system sizes $L=n\Delta$}{ Here, $\ell_c$ is held fixed ($\ell_c = L_{\text{max}}/\sqrt{A_{\text{max}}}\approx L_{\text{max}}/566$), such that the activity number $A= L^2/\ell_c^2$ increases with system size $L$. The wave number is rescaled by the largest system size $L_{\text{max}}$, such that the axis shows the mode number in the largest system ($n=2048$).}
\end{center}
\end{figure}

Because of energy conservation, the total energy transfer vanishes (\cref{eq integrated-transfer}). However, its spectrum could be non-zero so that energy is transferred between scales. In this section, we show that, in fact, the entire spectrum of energy transfer, \cref{eq power-transfer-spectrum}, vanishes. We show that based on symmetries. Because the equations of our model are symmetric with respect to rotation and reflection, the factor $\left\langle \tilde{\theta}_{\vec{k}}\,\tilde{\psi}_{\vec{q}-\vec{k}}\,\tilde{\theta}^*_{\vec{q}}\right\rangle $ can be a function only of $|\vec{q}|$, $|\vec{k}|$, and $\vec{q}\cdot \vec{k} = |\vec{q}||\vec{k}|\cos\alpha$. Therefore, this factor is an even function of the angle $\alpha$ between $\vec{q}$ and $\vec{k}$. In contrast, the factor $(\vec{k}\times\vec{q})\cdot\hat{z} = \pm |\vec{q}| |\vec{k}| \sin{\alpha}$ is an odd function of $\alpha$. Therefore, the integrand of \cref{eq power-transfer-spectrum} is an odd function of $\alpha$, and hence the integral vanishes. This means that the energy transfer spectrum \cref{eq power-transfer-spectrum} vanishes for all $\vec{q}$, $T(\vec{q})=0$, implying that there is no energy transfer between scales. We verified this result in our numerical calculations (\cref{Fig 3,Fig S2}).

\section*{Energy and power spectra} \label{definitions}

\subsection*{Frank elastic energy spectrum}

The angle-averaged spectrum of the Frank elastic free energy density, $F_n(q)$, is defined by
\begin{equation}
\frac{\langle F_n\rangle}{\mathcal{A}} \equiv \int_0^\infty F_n(q)\,\dd q.
\end{equation}
Using \cref{eq non-isotropic-Frank-spectrum}, and assuming 
isotropic correlations of the director field to integrate over the angle of the wave vector, we obtain
\begin{equation} \label{eq isotropic-spectra}
\langle F_n\rangle = \pi K L^4 \int_0^\infty q^3 \langle |\tilde{\theta}_{\vec{q}}|^2 \rangle \,\dd q,
\end{equation}
from where
\begin{equation} \label{eq Frank-energy-spectrum}
F_n(q) = \pi K L^2 q^3 \langle |\tilde{\theta}_{\vec{q}}|^2 \rangle.
\end{equation}
In dimensionless variables,
\begin{equation} \label{eq Frank-energy-spectrum-dimensionless}
F_n(q) = \pi q^3 \langle |\tilde{\theta}_{\vec{q}}|^2 \rangle.
\end{equation}

\subsection*{Power spectral densities}

Similarly, the four contributions to the power spectral density, $D_s(q)$, $D_r(q)$, $I(q)$, and $T(\vec{q})$, are defined by
\begin{multline}
\frac{\langle D_s\rangle}{\mathcal{A}} \equiv \int_0^\infty D_s(q)\,\dd q,\quad \frac{\langle D_r\rangle}{\mathcal{A}} \equiv \int_0^\infty D_r(q)\,\dd q,\\
\frac{\langle I\rangle}{\mathcal{A}} \equiv \int_0^\infty I(q)\,\dd q,\quad \frac{\langle T\rangle}{\mathcal{A}} \equiv \int_0^\infty T(q)\,\dd q.
\end{multline}
Hence, assuming isotropic correlations, the angle-averaged spectra read
\begin{multline} \label{eq power-spectra}
D_s(q) = 2\pi q D_s(\vec{q}),\qquad D_r(q) = 2\pi q D_r(\vec{q}),\\
I(q) = 2\pi q I(\vec{q}),\qquad T(q) = 2\pi q T(\vec{q}),
\end{multline}
with $D_s(\vec{q})$, $D_r(\vec{q})$, $I(\vec{q})$, and $T(\vec{q})$ given in \cref{eq power-spectra-contributions,eq power-transfer-spectrum}.

\subsection*{Kinetic energy spectrum}

The spectrum of the kinetic energy density per unit mass, $E(q)$, is defined by
\begin{equation}
\frac{\langle E\rangle}{\mathcal{A}} \equiv \int_0^\infty E(q)\,\dd q,
\end{equation}
with $E$ given by \cref{eq kinetic-energy-enstrophy}. Proceeding as before, in the continuum limit and for a state with isotropic correlations of the flow field, we obtain
\begin{equation}
E(q) = \pi L^2 q \langle |\tilde{\vec{v}}_{\vec{q}}|^2\rangle = \pi L^2 q^3 \langle |\tilde{\psi}_{\vec{q}}|^2\rangle.
\end{equation}
In dimensionless variables,
\begin{equation} \label{eq kinetic-energy-spectrum}
E(q) = \pi q^3 \langle |\tilde{\psi}_{\vec{q}}|^2\rangle.
\end{equation}

\subsection*{Enstrophy spectrum}

Respectively, the enstrophy spectrum, $\mathcal{E}(q)$, is defined by
\begin{equation}
\frac{\langle \mathcal{E}\rangle}{\mathcal{A}} \equiv \int_0^\infty \mathcal{E}(q)\,\dd q,
\end{equation}
where the enstrophy $\mathcal{E}$ is defined in \cref{eq kinetic-energy-enstrophy}
in terms of the vorticity field $\omega = -\nabla^2\psi$. In the same way as before, we obtain
\begin{equation}
\mathcal{E}(q) = 2\pi L^2 q \langle |\tilde{\omega}_{\vec{q}}|^2\rangle = 2\pi L^2 q^5  \langle |\tilde{\psi}_{\vec{q}}|^2\rangle.
\end{equation}
In dimensionless variables,
\begin{equation} \label{eq enstrophy-spectrum}
\mathcal{E}(q) = 2\pi q^5  \langle |\tilde{\psi}_{\vec{q}}|^2\rangle.
\end{equation}
Hence, the enstrophy spectrum is simply proportional to the shear dissipation spectrum, $D_s(q) = \eta\mathcal{E}(q)$, and it is also directly related to the kinetic energy spectrum, $\mathcal{E}(q) = 2 q^2 E(q)$.

\begin{figure}[tb]
\begin{center}
\includegraphics[width=0.75\columnwidth]{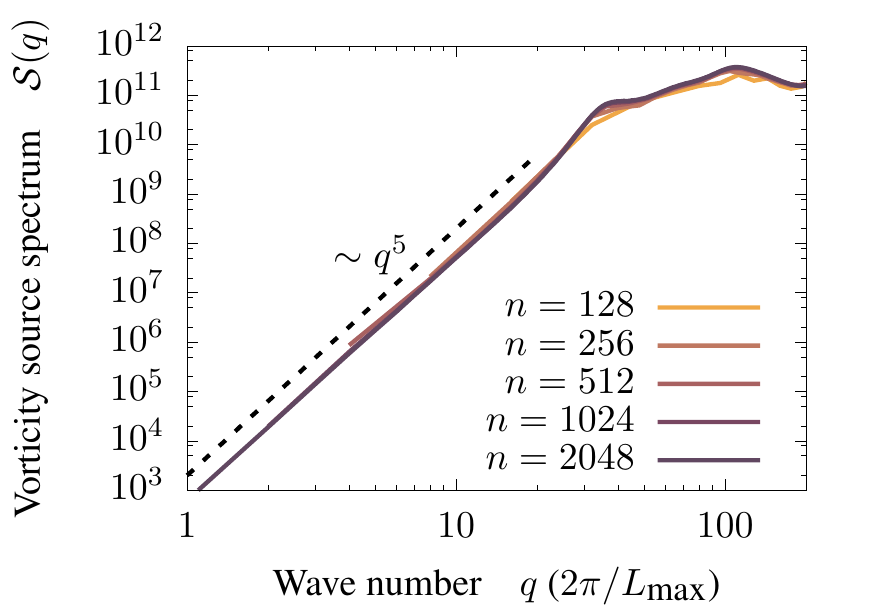}
\bfcaption{\label{Fig S3}Power spectrum of the vorticity source \cref{eq vorticity-source-spectrum} for different system sizes $L=n\Delta$}{ Here, $\ell_c$ is held fixed ($\ell_c = L_{\text{max}}/\sqrt{A_{\text{max}}}\approx L_{\text{max}}/566$), such that the activity number $A= L^2/\ell_c^2$ increases with system size $L$. The wave number is rescaled by the largest system size $L_{\text{max}}$, such that the axis shows the mode number in the largest system ($n=2048$).}
\end{center}
\end{figure}

\subsection*{Vorticity source power spectrum}

Finally, in \cref{eq biharmonic} we introduced the vorticity source $s(\vec{r},t)$. In dimensional form, it reads
\begin{equation}
s = \frac{K}{2\eta} \nabla^4 \theta + \frac{\zeta}{\eta} \left[\frac{1}{2}\left[\partial_x^2 - \partial_y^2\right] \sin 2\theta - \partial_{xy}^2 \cos 2\theta\right].
\end{equation}
Its power spectrum, $\mathcal{S}(q)$, is defined by
\begin{equation}
\frac{\langle \mathcal{S}\rangle}{\mathcal{A}} \equiv \int_0^\infty \mathcal{S}(q)\,\dd q,
\end{equation}
where we have introduced the \emph{vorticity source energy}
\begin{equation}
\mathcal{S} \equiv \int_{\mathcal{A}} s^2\,\dd^2\vec{r}.
\end{equation}
Analogously to the previous spectra, its spectrum reads
\begin{equation}
\mathcal{S}(q) = 2\pi L^2 q\langle |\tilde{s}_{\vec{q}}|^2\rangle.
\end{equation}
In dimensionless variables,
\begin{equation} \label{eq vorticity-source-spectrum}
\mathcal{S}(q) = 2\pi q\langle |\tilde{s}_{\vec{q}}|^2\rangle.
\end{equation}

\section*{Supplementary movies}

\subsection*{Supplementary movie 1} \label{movie1}

Evolution of the director angle $\theta(\vec{r},t)$ at activity number $A=500$, in degrees ($^\circ$). Angles $\theta= 0$ and $180^\circ$ correspond to the same orientation of the director field, and hence they are displayed in the same color. From the initial condition with a uniform director field, the system initially develops a stripe pattern, which then undergoes a zig-zag instability to form a vortex pattern as shown in \cref{Fig 1c}. This movie shows 40 snapshots of a simulation that runs for a time $t=0.16\tau_r$ on a grid of $256\times 256$ points.

\subsection*{Supplementary movie 2} \label{movie2}

Evolution of the director angle $\theta(\vec{r},t)$ at activity number $A=3.2\cdot 10^5$, in degrees ($^\circ$). Angles $\theta= 0$ and $180^\circ$ correspond to the same orientation of the director field, and hence they are displayed in the same color. As shown in \cref{Fig 2a}, at high activity, the system features a disordered pattern of small orientation domains with persistent dynamics indicative of spatiotemporal chaos. This movie shows 100 snapshots of a simulation that runs for a time $t=0.1\tau_r$ on a grid of $1024\times 1024$ points.

\subsection*{Supplementary movie 3} \label{movie3}

Evolution of the stream function $\psi(\vec{r},t)$ at activity number $A=2\cdot 10^4$, in units of $\tau_r^{-1} = K/(\gamma L^2)$. Local maxima and minima of the stream function correspond to counter-clockwise and clockwise flow circulation, respectively. As shown in \cref{Fig 4a}, at high activity, the system develops transient large patches of correlated flow. These patches encompass smaller coherent vortices, which are seen here as small ripples of the stream function. This movie shows 300 snapshots of a simulation that runs for a time $t=0.03\tau_r$ on a grid of $512\times 512$ points.

\end{document}